\documentclass[11pt,onecolumn,amssymb,nofootinbib]{revtex4}
\usepackage{amsmath, amsthm, amscd, amssymb}
\usepackage{bm}
\usepackage{bbm}

\begin{document}
\title{A frame-dependent gravitational effective action mimics a cosmological constant, but modifies the
black hole horizon}
\author{Stephen L. Adler}
\email{adler@ias.edu}
\affiliation{Institute for Advanced Study, Einstein Drive, Princeton, NJ 08540, USA}
\leftline{Essay written for the Gravity Research Foundation 2016 Awards for Essays on Gravitation.}
\leftline{Submitted Feb. 1, 2016}

\begin{abstract}
A frame dependent effective action motivated by the postulates of three-space general coordinate invariance and
Weyl scaling invariance exactly mimics a cosmological constant in Robertson-Walker spacetimes.  However, in a
static spherically symmetric Schwarzschild-like geometry it modifies the black hole horizon structure within microscopic distances
of the nominal horizon, in such a way that $g_{00}$ never vanishes.  This could have important implications
for the black hole ``information paradox''.
\end{abstract}
\maketitle

The experimental observation of an accelerated expansion of the universe has been interpreted as evidence for a cosmological
term in the gravitational action of the usual form
\begin{equation}\label{usual}
S_{\rm cosm}=-\frac{\Lambda}{8 \pi G} \int d^4x (^{(4)}g)^{1/2}~~~,
\end{equation}
with $\Lambda=3H_0^2\Omega_{\Lambda}$  in terms of the Hubble constant $H_0$ and the cosmological fraction $\Omega_{\Lambda}\simeq 0.72$.
This functional form incorporates the usual assumption that gravitational physics is four-space general coordinate invariant,
with no frame dependence in the fundamental action.   At the time when Einstein formulated General Relativity, this assumption
seemed entirely natural, since the relativity principle of special relativity dictates that all frames in uniform motion are
equivalent, with no possible way of singling out one uniformly moving frame as more special than the others.

The discovery of the cosmological microwave background (CMB) radiation, and the careful mapping of its temperature dependence
over the sky, undermines the assumption that it is impossible to specify a preferred uniformly moving frame of reference.
Consider an observer in an enclosed laboratory, in a state of uniform motion.  From observations within the laboratory, she cannot
detect this motion.   But if she drills a hole in the laboratory wall and erects a radiometer to measure the angular variation of
the CMB, she finds that the CMB provides a reference inertial frame.  By measuring the dipole component of the angular variation
of the CMB, she can infer her absolute velocity with respect to the CMB rest frame. Such a measurement shows that the solar system
is moving with a velocity of 369 km/s relative to the CMB frame.

Given that the CMB provides a preferred reference frame, it is then natural to ask whether there can be other physical effects
associated with this frame. In particular, could the assumption of a cosmological term with the frame-independent
form given in Eq. \eqref{usual} be replaced with a frame-dependent effective action term that has the same cosmological
consequences, but has different implications for gravitational physics in other contexts?

We begin by making the natural assumption that if there are preferred frame gravitational effects, they are rotationally
invariant in the CMB rest frame, since simplicity argues that there should be at most one preferred frame.  More generally,
we shall assume that any preferred frame effects are invariant under three-space general coordinate transformations, but
not under full four-space general coordinate transformations.  A significant additional restriction is obtained by
assuming invariance under global Weyl scaling transformations of the metric $g_{\mu\nu}$ by a general constant $\lambda$,
\begin{equation}\label{weyl}
g_{\mu\nu}(x) \to \lambda^2 g_{\mu\nu}(x)~,~~~
g^{\mu\nu}(x) \to \lambda^{-2} g^{\mu\nu}(x)~~~.
\end{equation}
Our introduction of this assumption was motivated in \cite{adler1} by reference to a pre-quantum trace dynamics theory \cite{adler2},
in which pre-quantum degrees of freedom are averaged over a canonical ensemble to give rise to quantum field theory as a thermodynamic
average.  Since the ensemble is constructed from
the trace Hamiltonian, it picks out a preferred frame, and for massless pre-quantum fields the ensemble can be shown to be invariant under Weyl scaling transformations of the metric and the pre-quantum fields.  Consequently, a residual gravitational effective action arising
from fluctuations in the pre-quantum fields will be invariant under the transformation of Eq. \eqref{weyl}.  But other motivations for
the assumption of Weyl scaling invariance can be given.  For example, in an Essay on Gravitation last year, 't Hooft \cite{thooft} argued
that physics should be invariant under local conformal symmetry, of which the Weyl scaling of Eq. \eqref{weyl} is a special case.

From the constraints of three-space general coordinate transformation invariance and Weyl scaling invariance,
to zeroth order in metric derivatives a residual preferred frame  gravitational effective action must have the form \cite{adler1}
\begin{equation}\label{effact1}
\Delta S_{\rm g}=\int d^4x (^{(4)}g)^{1/2}  (g_{00})^{-2} A(g_{0i}g_{0j}g^{ij}/g_{00},D^ig_{ij}D^j/g_{00},g_{0i}D^i/g_{00})~~~,
\end{equation}
with $D^i$ defined through the co-factor expansion of $^{(4)}g$ by $^{(4)}g/^{(3)}g=g_{00}+g_{0i}D^i$.
For the important special class of diagonal metrics for which $g_{0i}=D^i=0$, the effective action of Eq. \eqref{effact1} simplifies
to
\begin{equation}\label{effact2}
\Delta S_{\rm g}=A_0 \int d^4x (^{(4)}g)^{1/2} (g_{00})^{-2}~~~,
\end{equation}
with $A_0=A(0,0,0)$ a constant.

When particulate matter with action $S_{\rm pm}$ is present, the total action is the sum of the Einstein-Hilbert gravitational
action, the matter action $S_{\rm pm}$, and the effective action $\Delta S_{\rm g}$.  While the Einstein-Hilbert action and
 the matter action are four-space general coordinate invariant, the frame-dependent effective action $\Delta S_{\rm g}$ is
invariant only under the subset of general coordinate transformations that act on the spatial coordinates $\vec x$, while leaving
the time coordinate $t$ invariant.  Consequently, the stress-energy tensor obtained by varying $\Delta S_{\rm g}$ with respect
to the full metric $g_{\mu\nu}$  will not satisfy the covariant conservation condition, and thus cannot be used as a source
for the full spacetime Einstein equations.  However, it is consistent to include $\Delta S_{\rm g}$ in the source for the spatial
components of the Einstein tensor $G^{ij}$ in the preferred rest frame, giving the following rules:
\begin{enumerate}
\item  The spatial components of the Einstein equations are obtained by varying the full action with respect to $g_{ij}$, giving
\begin{equation}\label{einstein}
G^{ij}+8\pi G (\Delta T_{\rm g}^{ij} + T^{ij}_{\rm pm})=0  ~~~,
\end{equation}
with $T^{ij}_{\rm pm}$ the spatial components of the usual particulate matter stress-energy tensor, and with
$\Delta T_{\rm g}^{ij}$ given by
\begin{equation}\label{deltat}
\delta \Delta S_{\rm g }=-\frac{1}{2} \int d^4x (^{(4)}g)^{1/2} \Delta T_{\rm g}^{ij} \delta g_{ij}~~~.
\end{equation}
\item  The components of the Einstein tensor $G^{0i}=G^{i0}$ and $G^{00}$ are obtained from the
Bianchi identities with $G^{ij}$ as input, and from them we can infer the conserving extensions $\Delta T_{\rm g}^{i0}$ and
$\Delta T_{\rm g}^{00}$ of the spatial stress-energy tensor components  $\Delta T_{\rm g}^{ij}$.  Equivalently, we can infer these by imposing
the covariant conservation condition on the tensor $\Delta T_{\rm g}^{\mu \nu}$,  with $\Delta T_{\rm g}^{ij}$ as input.
\end{enumerate}

Let us now examine the implications of the frame-dependent effective action for two important spacetime geometries, the
cosmological Robertson-Walker line element, and the spherically symmetric  line element.  Both of these
have $g_{0i}=D^i=0$, and so we can use the simplified form of $\Delta S_{\rm g}$ in Eq. \eqref{effact2}.  Since the Roberston-Walker line
element has $g_{00}=1$, Eq. \eqref{effact2} simplifies to
\begin{equation}\label{effect3}
\Delta S_{\rm g}=A_0 \int d^4x (^{(4)}g)^{1/2} ~~~.
\end{equation}
Varying the spatial components of the metric, we find from Eq. \eqref{deltat} that
\begin{equation}\label{deltat1}
\Delta T_{\rm g}^{ij}=-A_0 g^{ij}~~~,
\end{equation}
for which the conserving extension is obviously given by
\begin{equation}\label{deltat2}
\Delta T_{\rm g}^{\mu\nu}=-A_0g^{\mu\nu}~~~.
\end{equation}
Thus, for a homogenous, isotropic Robertson-Walker cosmology, the frame dependent effective action has {\it exactly} the
structure of a cosmological constant. Hence observation of a nonzero cosmological fraction $\Omega_{\Lambda}$ does
not necessarily indicate the presence of a standard cosmological term  of Eq. \eqref{usual}, but instead could
indicate the presence of a frame-dependent effective action of Eqs. \eqref{effact1} and \eqref{effact2}, with
the constant $A_0$ given by
\begin{equation}\label{a0}
A_0=-\frac{\Lambda}{8\pi G}=-\frac{3 H_0^2 \Omega_{\Lambda} }{8\pi G}~~~.
\end{equation}
Note that in inferring the value of $A_0$ from the observed cosmological constant we have not given an explanation
of why $\Lambda$ is so small in Planck mass units.

Consider next the spherically symmetric line element
\begin{equation}\label{spher1}
g_{00}=B(r)~,~~G_{rr}=-A(r)~,~~g_{\theta\theta}=-r^2~,~~g_{\phi\phi}=-r^2\sin^2\theta~~~,
\end{equation}
for which $\Delta S_{\rm g}$ and $\Delta T_{g}^{ij}$ become
\begin{equation}\label{effactspher}
 \Delta S_{\rm g}=-\frac{\Lambda}{8\pi G} \int d^4x (^{(4)}g)^{1/2} B(r)^{-2}~,~~
\Delta T_{\rm g}^{ij}= \frac{\Lambda}{8\pi G} g^{ij}/B(r)^2~~~.
\end{equation}
Covariant conservation of $\Delta T_{\rm g}^{\mu\nu}$ then implies \cite{adler3} that $\Delta T_{\rm g}^{0i}=0$ and
\begin{equation}\label{spher3}
\Delta T_{\rm g}^{00}=-\frac{3\Lambda}{8 \pi G B(r)}~~~.
\end{equation}

A detailed numerical and analytic study of the static, spherically symmetric vacuum Einstein equations as modified by the effective action
$ \Delta S_{\rm g}$ of Eq. \eqref{effactspher} has been given by Adler and Ramazano\v glu \cite{adler3}.  The key results found there are:
\begin{enumerate}
\item  As might be anticipated from the factor $g_{00}^{-2}$ in $\Delta S_{\rm g}$, the existence of a horizon where $g_{00}$ vanishes is
suppressed.  Instead $g_{00}$ is nonvanishing for $0<r<\infty$. In spherical coordinates $g_{00}$ develops a square root branch point
at a finite value $r=a$, and is complex for $0<r<a$.  This branch point is a coordinate singularity, since it is absent in
isotropic coordinates, where again $g_{00}$ never vanishes.
\item  At macroscopic distances $>>10^{-17} {\cal M}$ cm outside the nominal horizon (where ${\cal M}$ is the black hole mass in
solar mass units) the numerical solutions closely approximate the standard Schwarzschild form until cosmological distances are
reached.  Within $10^{-17} {\cal M}$ cm from the nominal horizon, the behavior of $g_{00}$ is modified.
\item In contrast to  the Schwarzschild-de Sitter solution arising from a spherical geometry with a standard cosmological
constant given by $S_{\rm cosm}$, which has a non-vanishing curvature scalar $R \propto \Lambda$, the Schwarzschild-like solution arising from $\Delta S_{\rm g}$
has the curvature scalar  $R$ identically zero, just as does the usual vacuum Schwarzschild solution.
\item At cosmological distances the solution develops a physical singularity, which may be a reflection of the fact that a static metric
is too restrictive an Ansatz.   A numerical study relating to this is given in \cite{adler3}.
\end{enumerate}

To conclude, a frame-dependent modification of the gravitational action, as constrained by the requirements of three-space general
coordinate invariance and Weyl scaling invariance, mimics a standard cosmological constant for Robertson-Walker spacetimes.  But
the modified gravitational action
alters the horizon structure for Schwarzschild-like black hole solutions of the  Einstein equations.  Thus allowing  frame dependence
in the gravitational effective action may have significant implications for  black hole horizon physics, including the infamous ``information paradox'', suggesting an agenda for further study.

\end{document}